\documentclass[preprint2]{aastex}
\begin{document}
\title{Concerning the Distance to the Center of the Milky Way and its Structure}
\author{Daniel J. Majaess}
\affil{Saint Mary's University, Halifax, Nova Scotia, Canada}
\affil{The Abbey Ridge Observatory, Stillwater Lake, Nova Scotia, Canada}
\email{dmajaess@ap.smu.ca}

\begin{abstract}
The distance to the Galactic center inferred from OGLE RR Lyrae variables observed in the direction of the bulge is $R_0=8.1\pm0.6$ kpc.  An accurate determination of $R_0$ is hindered by countless effects that include an ambiguous extinction law, a bias for smaller values of $R_0$ because of a preferential sampling of variable stars toward the near side of the bulge owing to extinction, and an uncertainty in characterizing how a mean distance to the group of variable stars relates to $R_0$.  A $VI$-based period-reddening relation for RR Lyrae variables is derived to map extinction throughout the bulge. The reddening inferred from RR Lyrae variables in the Galactic bulge, LMC, SMC, and IC 1613 match that established from OGLE red clump giants and classical Cepheids.  RR Lyrae variables obey a period-colour ($VI$) relation that is relatively insensitive to metallicity.  Edge-on and face-on illustrations of the Milky Way are constructed by mapping the bulge RR Lyrae variables in tandem with cataloged red clump giants, globular clusters, planetary nebulae, classical Cepheids, young open clusters, HII regions, and molecular clouds.   The sample of RR Lyrae variables do not trace a prominent Galactic bar or triaxial bulge oriented at $\phi\simeq25 \degr$. 
\end{abstract} 

\section{Introduction}
Recent estimates of the distance to the center of Milky Way range from $R_0\simeq7-9$ kpc \citep{gb05,bi06,fe08,gr08,va09,ma09,mat09}.  The standard error associated with $R_0$ as inferred from variable stars is often smallest owing to sizeable statistics ($\le5\%$, se$=\sigma/\sqrt{n}$).  Yet it remains a challenge to identify and mitigate the dominant source of error, namely the systemic uncertainties.  In this study, several effects are discussed that conspire to inhibit an accurate determination of $R_0$ from the photometry of variable stars.  RR Lyrae variables detected in the direction of the bulge by OGLE are utilized to estimate the distance to the Galactic center, to map extinction throughout the region surveyed, and to assess the morphology of the Galaxy in harmony with red clump giants and pertinent tracers. 

\begin{figure*}[!t]
\begin{center}
\includegraphics[width=8.2cm]{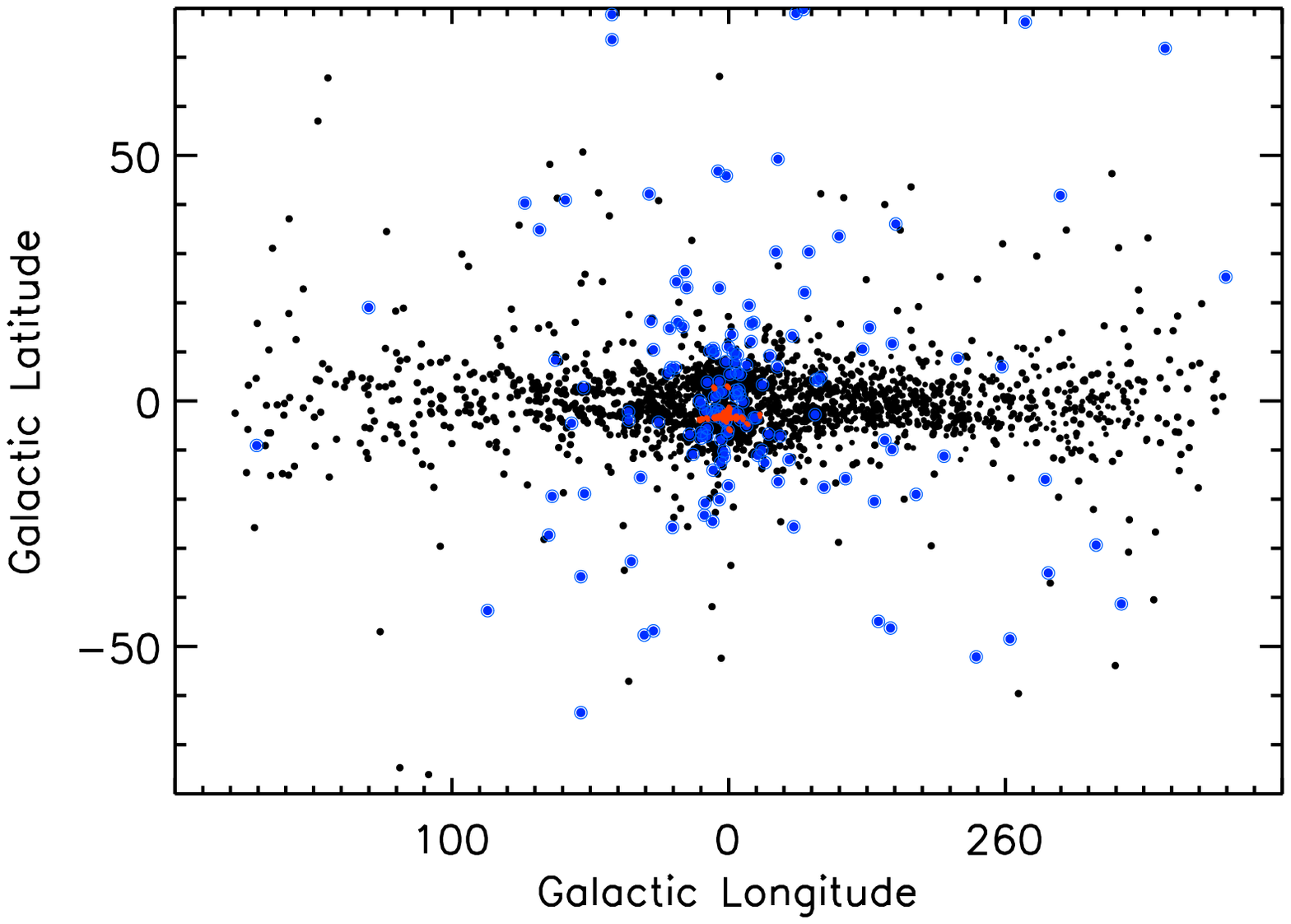} \\
\includegraphics[width=8.2cm]{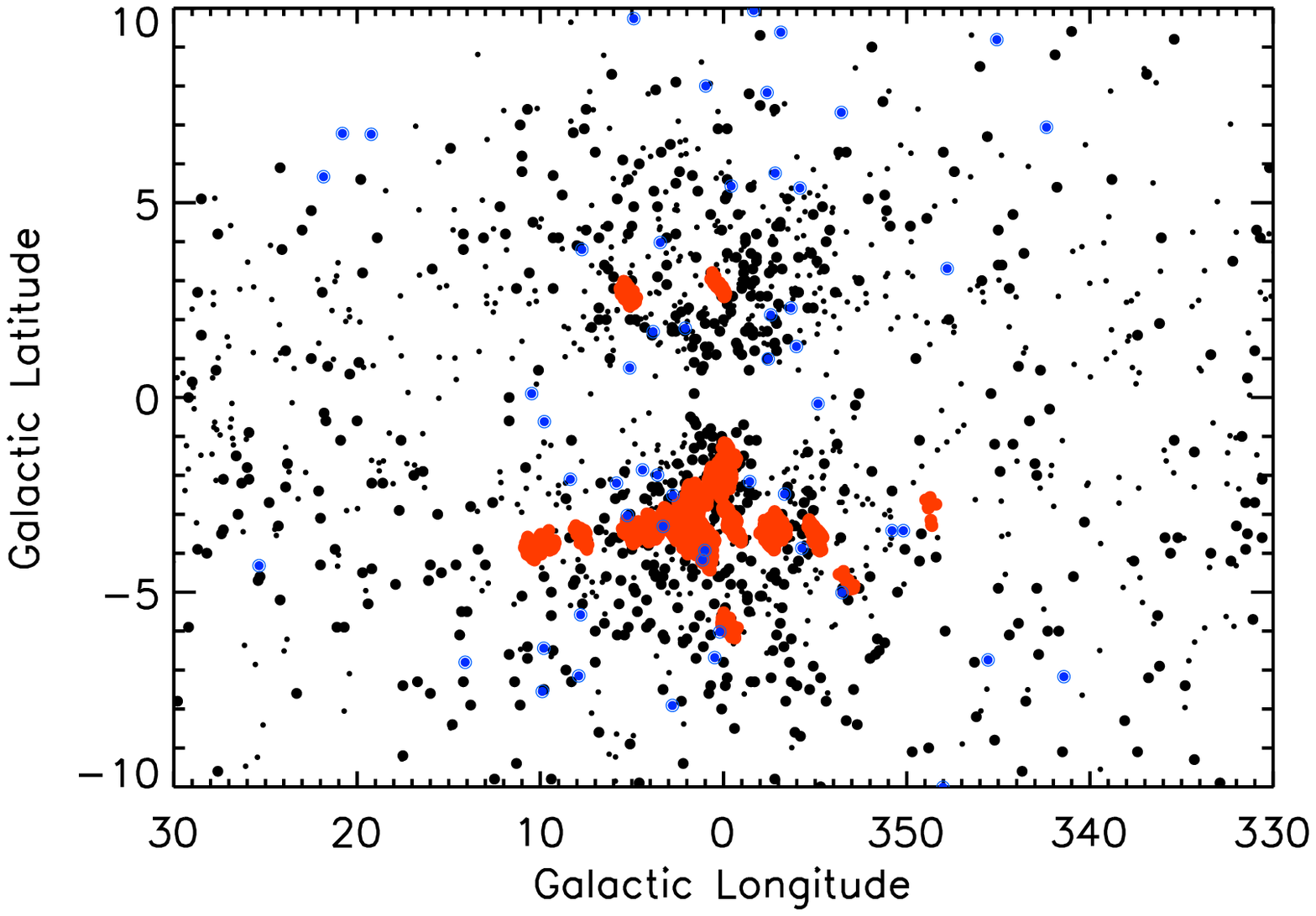}
\caption{\small{Edge-on view of the Milky Way as delineated by OGLE bulge RR Lyrae variables (red dots), planetary nebulae (black dots), and globular clusters (blue dots) in Galactic coordinate space.}} 
\label{fig1}
\end{center}
\end{figure*}

\section{Analysis}
\subsection{Sample \& Distribution}
The sample of RR Lyrae variables used here is that compiled by \citet*{co06} from the OGLE survey of Galactic bulge fields \citep{ud92,ud02,su04}.  Readers are referred to the comprehensive discussion in \citet{co06} regarding the construction of the database. Stars exhibiting spurious distances were not included in the present analysis.

\begin{figure*}[!t]
\begin{center}
\includegraphics[width=9cm]{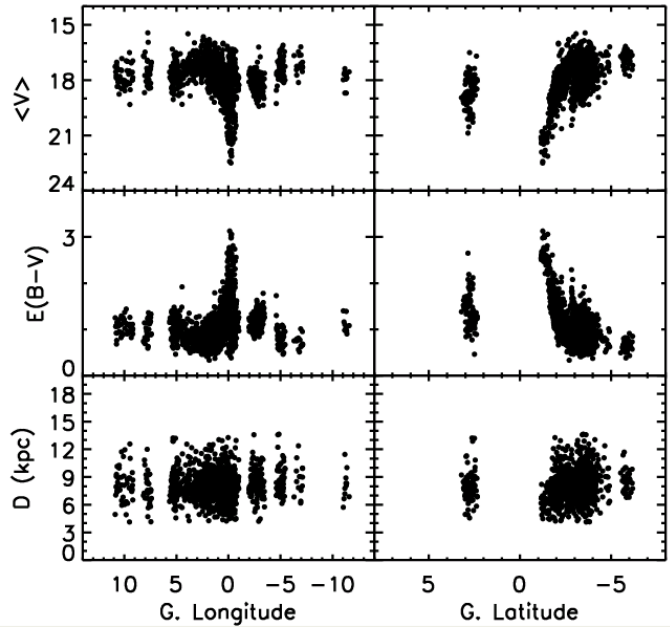}
\caption{\small{The mean visual magnitude, colour-excess, and distance associated with the RR Lyrae variables as a function of Galactic longitude and latitude.}}
\label{fig2}
\end{center}
\end{figure*}

It is instructive to begin by examining the distribution of RR Lyrae variables in position, magnitude, extinction, and distance space.  The locations of the variables are mapped on an edge-on model of our Galaxy as illustrated by planetary nebulae and globular clusters (Fig.~\ref{fig1}).  The distribution of planetary nebulae in Galactic coordinate space was compiled from the catalogs of \citet{kou01} and MASH I \& II \citep{pa06,mi08}. \citet{ha96} tabulated the relevant data for globular clusters.  Planetary nebulae, whose progenitors are primarily old low mass objects, outline the Galactic bulge where their distribution peaks rather clearly \citep[see Fig.~1 of][]{ma07}. RR Lyrae variables are not sampled in areas tied to anomalous extinction near the plane ($A_V\ge8$, Fig.~\ref{fig2}), and a similar absence is noted for planetary nebulae (Fig.~\ref{fig1}, right).  

The distribution of the sample's mean magnitude as a function of Galactic position indicates uneven sampling (Fig.~\ref{fig2}).  The survey proceeds deeper in tandem with the need to overcome increasing extinction toward the Galactic plane, namely $\langle V \rangle \simeq17$ at $b\simeq-6 \degr$, $\langle V \rangle \simeq19$ at $b\simeq3 \degr$, to $\langle V \rangle \simeq21$ at $b\simeq-1 \degr$.  

\subsection{Extinction}
\begin{figure}[!t]
\includegraphics[width=6.6cm]{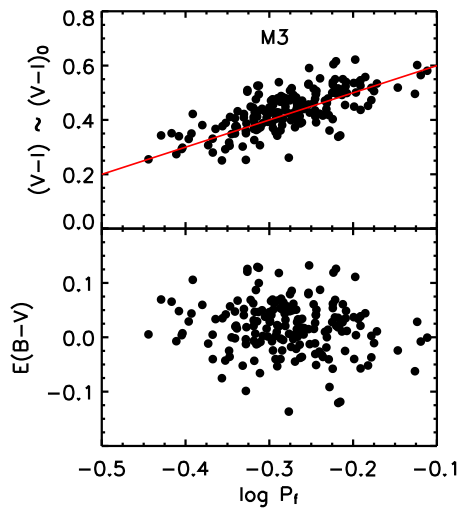}
\caption{\small{Top, $VI$ colours for RR Lyrae variables in M3 exhibit a period dependence \citep*[photometry from][]{ben06}.  The overplotted relation is Eqn. \ref{eqn1} (red line) or $(V-I)\simeq(V-I)_0\simeq 0.7+ \log{P_f}$. Bottom, the computed colour-excess using Eqn. \ref{eqn1}.}}
\label{fig8}
\end{figure}

Extinction throughout the bulge can be assessed by constructing a period-reddening relation for RR Lyrae variables, in similar fashion to classical Cepheids \citep{ma08,ma09c}.  That relation shall also facilitate the mapping of interstellar reddening for regions throughout the Milky Way and beyond. An RR Lyrae variable's colour excess may be \textit{approximated} by assuming that:
\begin{eqnarray}
\label{eqnred}
\nonumber
E_{B-V}\simeq \alpha \log{P} + \beta (m_{\lambda 1} - m_{\lambda 2}) + \phi
\end{eqnarray}
where $\alpha$, $\beta$, and $\phi$ are coefficients that can be derived by minimizing the $\chi^2$ statistic for a calibrating data set, and $m_{\lambda 1}$ and $m_{\lambda 2}$ are mean photometric magnitudes in different passbands.  The calibrators are RR Lyrae variables in the globular clusters M3 \citep{ha05}, M54 \citep{ls00}, M92 \citep{ko01}, and NGC 6441 \citep{pr03}.  Reddenings for the calibrating globular clusters were acquired from various studies \citep[e.g.,][]{ha96,ma09c}. The optimum solution is:
\begin{equation}
\label{eqn1}
E_{B-V}\simeq -0.88 \log{P_f} + 0.87 (V-I) - 0.61
\end{equation}
which reproduces the calibrating set with an average uncertainty of $\pm0.03$ magnitude. The true scatter applying to use of the relationship for individual RR Lyrae variables may be larger, particularly for stars near the edge of the instability strip.  The relation can provide first order estimates to complement space reddenings \citep{be02,be07,lc07,tu09}.  RR Lyrae variables pulsating in the overtone were shifted by $\log{P_f}\simeq\log{P_o}+0.13$ to yield the equivalent fundamental mode period \citep{wn96,so03,gr07,hu09}.   Alternatively, M3 offers a unique opportunity to infer the intrinsic $VI$ colours of RR Lyrae variables directly since foreground extinction along the globular cluster's line of sight is negligible \citep{mr69}.  Thus a formal fit to $VI$ photometry of RR Lyrae variables in M3 may be employed to establish reddenings.  An interpretation of \citet{ben06} M3 photometry is given in Fig.~\ref{fig8}.

\begin{figure}[!t]
\includegraphics[width=6.6cm]{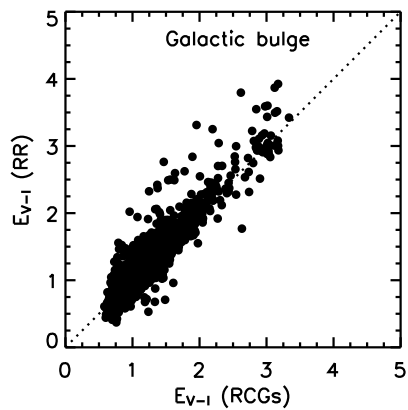}
\caption{\small{The colour-excess inferred from bulge RR Lyrae variables (Eqn.~\ref{eqn1}) match estimates compiled by \citet{su04} for adjacent red clump giants.}}
\label{fig4}
\end{figure}

Reddenings for locations sampled by the survey are mapped as a function of Galactic position (Fig.~\ref{fig2}).  The following general trend can be inferred regarding extinction throughout the bulge, namely that it is not symmetric or uniform.  For example, across $b\simeq3\degr$ the reddening varies from approximately  $E_{B-V}\simeq 0.6 \rightarrow 2.3$ (Fig.~\ref{fig2}).  The reddening throughout the entire sample ranges from $E_{B-V} \simeq 0.4 \rightarrow3.4$, with extinction increasing to a maximum near the dust ridden Galactic plane (Fig.~\ref{fig2}).  The estimated colour-excess is in satisfactory agreement with that inferred by \citet{su04} from adjacent OGLE red clump giants (Fig.~\ref{fig4}).  

\begin{deluxetable}{lccccc}
\tabletypesize{\scriptsize}
\tablecaption{Mean reddenings for the galaxies ($E_{B-V}$) \label{reds}} 
\tablewidth{0pt}
\tablehead{\colhead{Object} &\colhead{Classical Cepheids} &\colhead{RR Lyrae Var.} &\colhead{Red Clump Giants} &\colhead{\citealt{za04}} &\colhead{Photometry} \\ & & & \citep{ud99} & \citep{ng09} &}
\startdata
LMC & $0.14$ & $0.12$ & $0.14$ & $0.13$ & (1,2,6,7,8) \\
SMC & $0.13$ & $0.10$ & $0.09$ & - & (1,3) \\
IC1613 & $0.05$ & $0.05$ & - & - & (4,5)
\enddata
\tablecomments{(1) \citet{ud99}, (2) \citet{so09}, (3) \citet{so02}, (4) \citet{do01}, (5) \citet{ud01}, (6) \citet{so08b}, (7) \citet{ci00}, (8) \citet{ma02}.  The classical Cepheid reddenings are inferred from the Galactic calibration of \citet{ma09c}.  \textbf{***} The colour excess varies with position across the Magellanic Clouds and shall be elaborated upon elsewhere.}
\end{deluxetable}

The robustness of equation~\ref{eqn1} may be further evaluated.  Drawing upon photometry for RR Lyrae variables in the LMC \citep{ud98,so03,so09}, SMC \citep{ud98,so02}, and IC 1613 \citep{do01}, the resulting mean reddenings for the galaxies are consistent with estimates inferred from red clump giants, a Galactic classical Cepheid relation, and other means (Table~\ref{reds}).  That indicates RR Lyrae variables adhere to a period-colour ($VI$) relation which is relatively insensitive to metallicity.  Perhaps an expected result granted the slope of near infrared RR Lyrae variable and Cepheid distance relations are comparatively unaffected by chemical composition \citep{lo90,ud01,bo03b,pi04,pe04,so06,de06,be07,vl07,fo07,mat06,ma08,ma09,ma09c,ma09d}.  By contrast, readers should exhibit caution when employing $BV$ relations for Cepheid and RR Lyrae variables of differing abundance \citep{cc85,mf91,ch93,ta03,di07,ma08,ma09c}.  The computed colour-excess of the brightest member of the variable class, RR Lyrae, is $E_{B-V}\simeq0.01$.  That agrees with both a value cited by \citet{fe08} and a field reddening inferred from 2MASS photometry using methods tested elsewhere \citep{bo04,bo06,ma07,ma08,bb09,tu09b,tu09d,tu09}.  The implied absolute magnitude for RR Lyrae is $M_V\simeq0.55$, assuming $E_{B-V}\simeq0.01$ (eqn.~\ref{eqn1}) and $d\simeq260$ pc \citep{be02,bon02,ma09d}. To establish the parameters $VI$ photometry from \textit{The Amateur Sky Survey} for RR Lyrae was utilized \citep{dr06}, although concerns persist regarding the survey's photometric zeropoint and the star's modulating amplitude.  An ephemeris to phase the $V$ \& $I$ data was adopted from the GEOS RR Lyr database \citep{bo02,lb04,lb07}.  

\begin{figure}[!t]
\includegraphics[width=7cm]{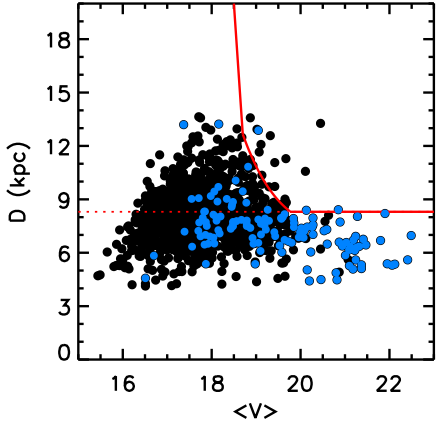}
\caption{\small{OGLE RR Lyrae variables nearest to the Galactic plane ($b\simeq-1.5\degr$) and fainter than $V\simeq19$ are not sample beyond $d\simeq8$ kpc (blue dots).}}
\label{fig3}
\end{figure}

\subsection{Distance}
A Type II Cepheid $VI$ reddening-free relation is employed to provide distances for the sample of RR Lyrae variables \citep[Eqn.~2,][]{ma09}.  Relying exclusively on the photometric surveys of fellow researchers, \citet{ma09d} reaffirmed that to \textit{first order} SX Phe, RR Lyrae variables, and Type II Cepheids may adhere to a common $VI$ period-magnitude-colour relation \citep[see also the interesting $JHK_s$ results of][]{mat06}.  A $VI$ Wesenheit diagram and function illustrate the underlying period-magnitude continuity unifying variables of the population II instability strip \citep{ma09d}, although perhaps somewhat imperfectly.  Admittedly, small statistics inhibit an elaborate analysis toward the SX Phe regime and RR Lyrae variables may exhibit a steeper Wesenheit function than Type II Cepheids.  However, distances inferred from Eqn.~2 of \citet{ma09} are, however, largely insensitive to that latter putative difference (Fig.~\ref{fig11}, see also \citealt{ma09d}).  The slopes of the Wesenheit functions characterizing shorter-period Type II Cepheids and RR Lyrae variables are generally consistent with the predictions of Marconi, Di Criscienzo, \& collaborators, underscoring the viability of their research team's pulsation models.  Readers are referred to studies by \citet{vb68}, \citet{ma82}, \citet{op83}, \citet{kj97,kw01}, \citet*{di04}, \citet{mf91,mf09}, and \citet{tu09} for a broader discussion regarding RR Lyrae and Cepheid Wesenheit relations.  The calibrators of the \citet{ma09} $VI$ distance relation were OGLE LMC Type II Cepheids \citep{ud99,so08}, with an adopted zeropoint to the LMC established from classical Cepheids and other means \citep[$(m-M)_0 \sim 18.50$,][]{ls94,gi00,fr01,be02,ma08,ma09d}.   The classical Cepheid zeropoint to the LMC was inferred from the photometry of \citet{ud99} and \citet{se02}, using the reddening-free distance relation of \citet{ma08}.  That relation is tied to a restricted subsample of Galactic cluster Cepheids \citep[e.g.,][]{tu02} and Cepheids with new HST parallaxes \citep{be07}.  

The mean distance to the sample of RR Lyrae variables observed in the direction of the bulge is $8.1\pm0.6$ kpc ($\sigma/2$).  Yet what is the relation to $R_0$, the distance to the Galactic center?  The mean is $R_0$ if the sampling were uniform across a symmetric bulge.  However, a fraction of RR Lyrae variables are not sampled equally at the rear and forefront of the bulge.  Indeed, as the survey nears the Galactic plane, the effects of extinction increase the magnitude threshold needed to adequately sample the bulge beyond the limit of the survey (Fig.~\ref{fig2},~\ref{fig3}).  The mean distance inferred from RR Lyrae variables near $b\simeq-1 \degr$ is $d\simeq7$ kpc, while at larger Galactic latitudes it is approximately a kiloparsec further (Fig.~\ref{fig2},~\ref{fig3}).   RR Lyrae variables fainter than $V\simeq19$ and nearest to the Galactic plane are not observed beyond $\simeq8$ kpc (Fig.~\ref{fig3}).  Establishing $R_0$ from that limited subsample shall yield a result systemically too close.  The data were therefore excluded from the derivation of $R_0$.  \citet{ma09} suggested the impact of the bias may be assessed by ascertaining $R_0$ via an alternative approach, namely by adding an estimate for the radius of the bulge to the distance to its near side.  Admittedly, that approach introduces new uncertainties and is idealistic granted the bulge may be triaxial (Fig.~\ref{fig5}). 

Establishing the distance to the Galactic center from either bulge Type II Cepheids \citep{ku03,ma09} or RR Lyrae variables yields an analogous result ($R_0\simeq8$ kpc).  However, that result depends rather sensitively on the extinction law adopted, particularly since the reddenings are inherently large (Fig.~\ref{fig2}).  The nature of the extinction law toward the bulge is actively debated and is thus a primary source of uncertainty in the determination of $R_0$ \citep{go01,ud03,ru04,su04,kc08}.   The \citet{ma09} distance relation employed here assumes an extinction law similar to that cited for bulge stars by \citet{ud03}.  In hindsight, however, perhaps the least squares approach adopted by \citet{ma08,ma09} to obtain the coefficients of the Cepheid (Type I \& II) distance relations, which includes an estimate for the pseudo extinction law term, should be forgone in favour of relations derived assuming an extinction law \textit{a priori}.  The matter shall be elaborated upon in a separate study.  Nevertheless, the distance derived here to the center of the Milky Way agrees with that inferred by geometric means \citep[$R_0\simeq7.6-8.3$ kpc,][Table 1]{va09}.  Note, however, that geometric-based estimates of $R_0$ exhibit scatter and non-zero uncertainties, important details which are often overlooked including previously by the author \citep{ma09}.

\subsection{Galactic Structure}
\label{sgstruc}
\begin{figure}[!t]
\includegraphics[width=6.5cm]{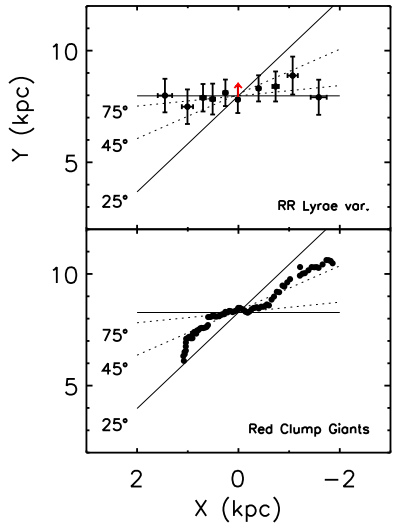}
\begin{center}
\includegraphics[width=4.5cm]{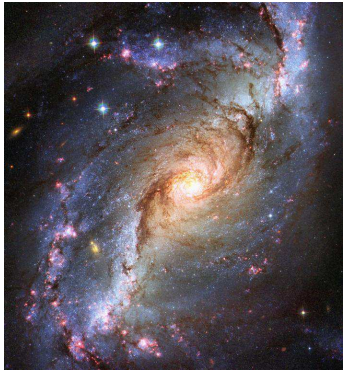}
\end{center}
\caption{\small{Top, an orientation of $\phi \simeq 25\degr$ (sun-gc line) is not indicative of RR Lyrae variables detected in the direction of the bulge.  Middle, red clump giants may trace a nested nuclear structure within a primary bar \citep[data from][]{ni05,ni06}.  Bottom, a reprocessed cropped portion of the HST image of NGC 1672.}}
\label{fig5}
\end{figure}

One commonly proposed scenario has the Milky Way exhibiting a bar oriented at $\phi\simeq25 \degr$ along the sun-Galactic center line.  The reputed bar typically extends from $-10\le \ell \le 10 \degr$, being nearer to the Sun for positive $\ell$. However, RR Lyrae variables do not appear to delineate a prominent bar or triaxial bulge at $\phi \simeq25 \degr$ (Fig.~\ref{fig2},~\ref{fig5}).  Incidentally, nor is extinction in the region of $5\le \ell \le 10\degr$ anomalous, as otherwise expected if observing through a dense thick bar (Fig.~\ref{fig2},~\ref{fig5}).   A formal fit to the variable star sample binned in $\ell$ yields $\phi \simeq77 \pm15 \degr$, which is also in agreement with an axisymmetric distribution (Fig.~\ref{fig5}, top).  \citet{al98} and \citet{ku08} likewise note that there is marginal evidence for a bar in the distribution of bulge RR Lyrae variables.  Readers are referred to their studies for a comprehensive discussion of the proposed rationale. 

\begin{figure}[!t]
\includegraphics[width=6.5cm]{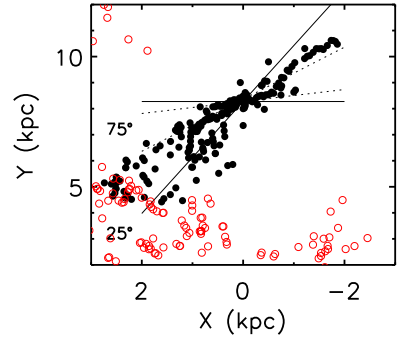}
\caption{\small{A compilation of results from studies of red clump giants exhibits unsatisfactory scatter.  At increasing distance from the Galactic center the red clump giants (black dots) may be sampling spiral features in the young disk.  HII region are symbolized by red open circles. The upcoming VVV survey shall bolster statistics in the fourth quadrant \citep{mi10}.}}
\label{fig10}
\end{figure}

\citet{ni05,ni06} infrared survey of red clump giants put forth detailed evidence that the Milky Way exhibits a distinct nuclear structure nested within a primary bar (Fig.~\ref{fig5}, middle). Indeed, considering the inner region of the galaxy NGC 1672 as a template (Fig.~\ref{fig5}, bottom), the \citeauthor{ni05} data may imply the presence of two minor pseudo spiral arms or spurs (note the dust lanes) that emanate from a nuclear structure and then reconnect to an underlying primary bar.  The nested feature is oriented at an angle ($\phi_{N}\simeq70\degr$) consistent with the distribution of most RR Lyrae variables (Fig.~\ref{fig5}).  RR Lyrae variables may in part sample that region which is nearly axisymmetric.  While the pseudo minor spiral arms connect to a primary bar at $\phi_{B}\sim30 \degr$.  The angle $\phi$ as often cited in the literature may be a mischaracterization or average of two (or multiple) distinct features (Fig.~\ref{fig5}, middle).  There lacks consensus on defining how that parameter should be ascertained from the data (Fig.~\ref{fig5}, middle --- e.g., a bulk mean, a mean from tip to tip, a mean for each structure $\phi_{N}$ \& $\phi_{B}$ (preferred), etc.). The observations of \citeauthor{ni05} are based on high resolution sampling: $8 \arcmin$ intervals from $-10\ge\ell\ge10\degr$ at $b\simeq1 \degr$.  By contrast, the data in Fig.~\ref{fig5} (top) consist of an admixture of RR Lyrae variables at differing $b$ (Fig.~\ref{fig1},~\ref{fig2}).  That complicates an interpretation of the RR Lyrae distribution since a degeneracy emerges owing to a correlation between distance and Galactic latitude ($b$, Fig.~\ref{fig2}).  For example, the most distant and deviant point in Fig.~\ref{fig5} (top) is inferred from high latitude bulge variables observed through low extinction ($b\simeq-5 \degr$, Fig.~\ref{fig2}).   Furthermore, \citet{su04} remarked that the ratio of total to selective extinction may vary weakly as a function of galactic longitude, increasing from positive to negative $\ell$.  The effect on distances becomes magnified granted the reddenings are inherently large toward the Galactic bulge and near the plane.  Adopting a mean $R_{\lambda}$ may tend to produce nearer distances for objects at positive $\ell$ relative to objects at negative $\ell$.  That follows the orientation of the proposed bar.  Admittedly, if that bias is real it affects the results derived here for the distribution of bulge RR Lyrae variables. 

\begin{figure*}[!t]
\begin{center}
\includegraphics[width=15cm]{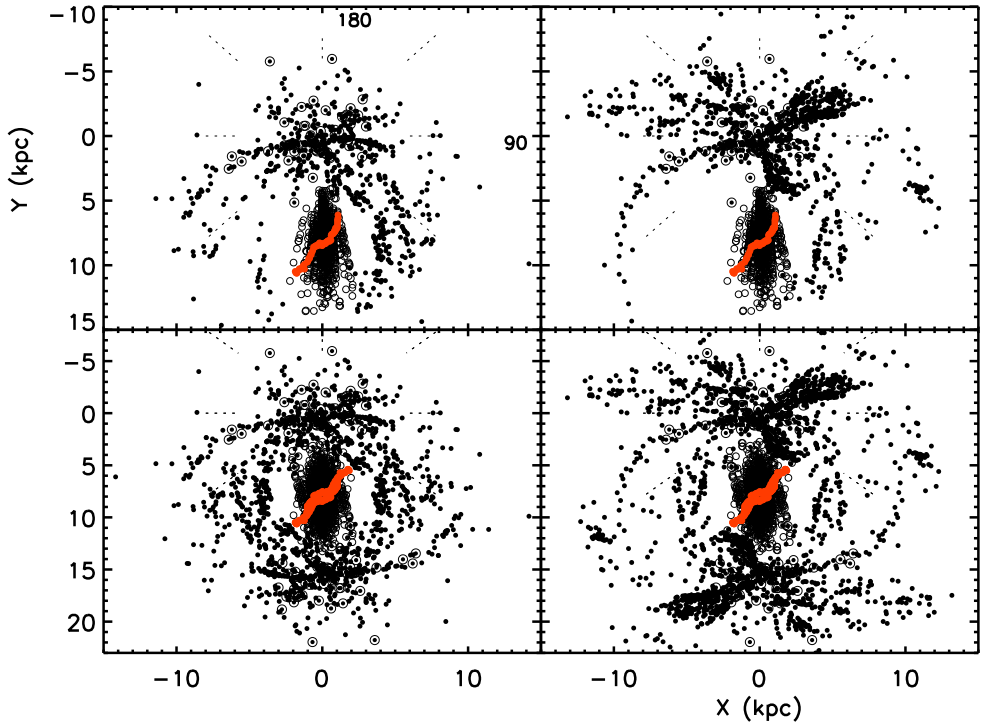}
\caption{\small{Face-on view of the Milky Way as delineated by bulge RR Lyrae variables and red clump giants (red), classical Cepheids, young open clusters, HII regions, and molecular clouds. Left, illustration without molecular clouds.  Right, illustration without HII regions.  \textit{Bottom, data flipped and mirrored to provide perspective}.}}
\label{fig7}
\end{center}
\end{figure*}

The red clump giant observations of \citeauthor{ni05} were adopted verbatim and yet the seminal nature of the implied result demands rigorous scrutiny.  Indeed, the reader should know there are countless concerns, as with all distance candles. For example, a compilation of results for red clump giants observed toward the Galactic bulge exhibits unsatisfactory scatter (Fig.~\ref{fig10}).  The scatter at increasing distance from the Galactic center is exacerbated from sampling spiral features in the young disk (Fig.~\ref{fig10}).  A separate overview of the conclusions from various red clump giant studies is given in \S 2.2 \& Table 1 of \citet{va09}.  The interpretation and evidence presented by \citet{st94}, \citet{ra07}, and \citet{cl08} should likewise be considered.

A face-on perspective of the Milky Way (Fig.~\ref{fig7}) is now assembled from cataloged RR Lyrae variables \citep{co06}, red clump giants \citep{ni05,ni06}, classical Cepheids \citep[e.g.,][]{sz77,sz80,sz81,be00}, young open clusters \citep{di02,mp03}, HII regions \& molecular clouds \citep{ho09}.  Classical Cepheids, young open clusters, HII regions, and molecular clouds trace the Milky Way's younger spiral arms \citep{wa58,bo59,ks63,ta70,op88,ef97,ma09,ma09b}.   An overdensity of HII regions and molecular clouds is observed near the interaction between the reputed bar and young disk [X,Y$\simeq$1.5,4 kpc] (Fig.~\ref{fig7}).  Interestingly, the Sagittarius-Carina arm may emanate from that region since it in part stems or branches from $\ell \le 35 \degr$ rather than $\ell \simeq 50 \degr$ (\citealt{ma09,ma09b}, see also Fig.~5 in \citealt{rus03}).  Superposed logarithmic spiral patterns ineptly characterize distinct features near the Sun, particularly segments of the putative Orion spur or Local and Sagittarius-Carina arms \citep{rus03,ma09,ma09b,ho09}.  Added flexibility is needed to consider spurs, and spiral arms that merge, branch, twist unexpectedly, and exhibit a degree of flocculence.  Such features are common amongst a sizeable fraction of the Universe's galaxies, while perfect grand-design spiral patterns are less prevalent (browse the Atlas of Galaxies or Galaxy Zoo Project, \citealt{sb88,ra09}).  Indeed, the commonly espoused scenario of the Sun within a spur indicates that such features are likely not unique, and exist elsewhere throughout the Galaxy.  More work is needed here, and a holistic approach that integrates RR Lyrae variable and red clump giant populations into analyses of the Galaxy's overall structure may facilitate an interpretation.  

\begin{figure*}[!t]
\includegraphics[width=17.5cm]{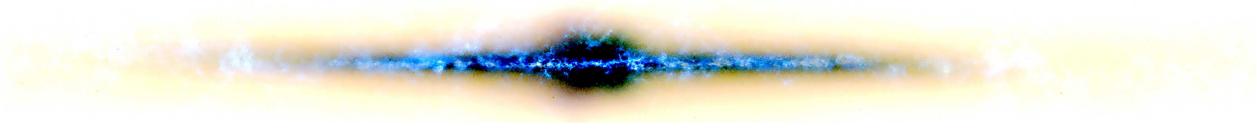}
\caption{\small{A reprocessed cropped portion of the 2MASS mosaic of the Milky Way \citep{cu03}.  The Galactic bulge exhibits a peanut-like morphology.}}
\label{fig6}
\end{figure*}

Complementing the edge-on illustration of the Milky Way displayed earlier (Fig.~\ref{fig1}), 2MASS IR observations imply that the Milky Way exhibits a peanut shaped bulge \citep[Fig.~\ref{fig6}, see also][]{we94}. That profile is argued by fellow researchers to indicate a bar seen edge-on \citep[][and references therein]{cb04}.  Conversely, the morphology of the Galactic bulge appears somewhat spherical in optical images \citep{ga60,bt09,me09}.  However, the bulge assumes a peanut-like geometry once anomalous extinction across the Aquila Rift is accounted for \citep[][and references therein]{st03,ma09b}.  Readers are encouraged to correlate CO markers tied to molecular complexes in Aquila and Lupus with the corresponding dark rifts in A. Mellinger's photographic mosaic of the Galaxy \citep[see Fig.~6 of][]{da01}.

\section{Summary \& Future Research}
In this study, RR Lyrae variables cataloged by \citet*{co06} from the OGLE survey were used to determine the distance to the Galactic center, to map extinction throughout the sample, and to facilitate an interpretation of the Milky Way's structure.

The implied distance to the center of the Galaxy is $R_0=8.1\pm0.6$ kpc.  An accurate determination of $R_0$ is hindered by countless sources.  It is insightful to examine a sample's distribution in position, magnitude, and extinction space, to assess how the mean distance to a group of variable stars detected in the direction of the Galactic bulge relates to $R_0$ (Fig.~\ref{fig1},~\ref{fig2},~\ref{fig3}).  Extinction imposes a preferential sampling of stars toward the near side of the bulge.  Consequently, a mean distance inferred from that restricted subsample shall promote smaller values of $R_0$. The effect is particularly acute for RR Lyrae variables near the Galactic plane ($b\simeq0\degr$, Fig.~\ref{fig2}), where excessive extinction increases the magnitude threshold needed to adequately sample the bulge beyond the limit of the survey (Fig.~\ref{fig2},~\ref{fig3}).  Furthermore, the supposed presence of a Galactic bar may bias estimates of $R_0$ depending on which bulge region(s) are sampled.  $R_0$ shall be systemically too large if inferred solely from bulge stars at negative $\ell$ that may outline the far side of the reputed Galactic bar (Fig.~\ref{fig5}).  Caution is warranted when ascertaining $R_0$ from groups of stars exhibiting poor statistics and sampling limited regions of the bulge.   Additional concerns persist regarding an ambiguous extinction law for bulge stars \citep[\textit{important},][]{go01,ud03,ru04,su04,kc08}, an uncertainty in the LMC's zeropoint which is implicitly tied to the $VI$-based reddening-free distance relations employed here \citep{gi00,fr01,be02,ta03}, an ongoing debate surrounding the contested effects of metallicity for Cepheid and RR Lyrae variables \citep{ud01,fr01,fe03,sm04,mot04,pi04,ro05,so06,ma06,bo08,sc09,ro08,ma08,ca09,ma09,ma09c,ma09d}, the effects of photometric contamination (e.g., blending \& crowding) on distances computed to variable stars \citep{su99,mo00,mo01,mo02,ma01,fr01,vi07,sm07,ma09c}, and floating photometric zeropoints owing to the difficulties in achieving standardization, particularly across a range in colour \citep[e.g.,][]{tu90,st04,sah06,jo08}.  The author suggests the evidence indicates that $VI$-based RR Lyrae and Cepheid distance and period-colour relations are relatively insensitive to metallcity, and thus by consequence, that the distance offset observed between metal-rich and metal-poor classical Cepheids occupying the inner and less crowded outer regions of remote galaxies arises primarily from other source(s) \citep[][see also \citealt{ud01,pi04,bo08}]{ma09c,ma09d}.  Readers are encouraged to also consider the dissenting views and varied interpretations presented in the works cited earlier.  Firm constraints on the effects of metallicity, and hence crowding and blending, may arise from a direct comparison of RR Lyrae variables, Type II Cepheids, and classical Cepheids at a common and comparitively nearby zeropoint (e.g., SMC, IC 1613).

The sample of RR Lyrae variables do not trace the signatures of a prominent bar or triaxial bulge oriented at $\phi\simeq 25 \degr$ (Fig.~\ref{fig2},~\ref{fig5}), as noted previously \citep[][and references therein]{al98}.  The stars exhibit a more axisymmetric distribution and may outline, in part, a nuclear structure (Fig.~\ref{fig5}).  A confident interpretation is complicated by the admixture of RR Lyrae variables at varying galactic positions.  By contrast, younger red clump giants may delineate a nearly axisymmetric nuclear structure ($\phi_{N}\simeq70\degr$, Fig.~\ref{fig5}) nested within a primary Galactic bar ($\phi_{B}\sim30\degr$).  Yet there are pertinent concerns with the aforementioned interpretation, and that found in the literature (\S \ref{sgstruc}, Fig.~\ref{fig5} \&~\ref{fig10}).  First, a compilation of results from several studies on red clump giants exhibits considerable scatter (Fig.~\ref{fig10}).  The scatter at increasing distance from the center of the Milky Way arises partly from sampling spiral features in the young disk (Fig.~\ref{fig10}).  Third, $\phi$ as currently cited in the literature may be a mischaracterization or average of two (or multiple) distinct features (e.g., $\phi_{N}$ \& $\phi_{B}$, Fig.~\ref{fig5}).  The structure of the Galaxy's inner region may be too complex to be ascribed a single linear term or angle $\phi$ (Fig.~\ref{fig5}).  Lastly, curiously, extinction in the region of $5\le \ell \le 10\degr$ as inferred from RR Lyrae variables is not anomalous, as otherwise expected if observing through a dense thick bar delineated by red clump giants (Fig.~\ref{fig2},~\ref{fig5}). 

Edge-on and face-on illustrations of the Milky Way are constructed by mapping the OGLE RR Lyrae variable sample in tandem with cataloged red clump giants, planetary nebulae, globular clusters, classical Cepheids, young open clusters, HII regions, and molecular clouds (Fig.~\ref{fig1},~\ref{fig7}).   An abundance of HII regions and molecular clouds are observed near the boundary between the reputed Galactic bar and young disk [X,Y$\simeq$1.5,4 kpc] (Fig.~\ref{fig7}).  Moreover, the Sagittarius-Carina spiral arm may in part originally stem or branch from near that region (Fig.~\ref{fig7}).  

\begin{figure}[!t]
\includegraphics[width=6.7cm]{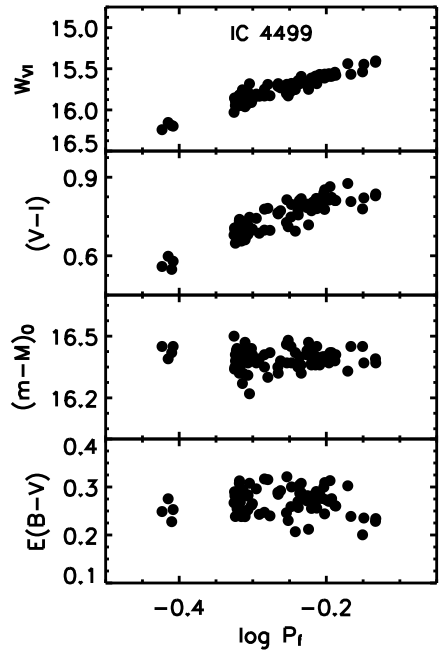}
\caption{\small{RR Lyrae variables follow $VI$ period-colour and Wesenheit period-magnitude-colour relations (e.g., RRe$\rightarrow$RRab variables in the globular cluster IC 4499, photometry from \citealt{wn96}).  The cluster's distance and mean colour-excess are $(m-M)_0=16.40\pm0.04$ (sd) and $E_{B-V}=0.27\pm0.03$ (sd).  Note the minimal \textit{internal} scatter.}}
\label{fig11}
\end{figure}

 A $VI$-based RR Lyrae period-reddening relation derived here reaffirms that extinction throughout the bulge is highly inhomogeneous, varying from $E_{B-V} \simeq 0.4 \rightarrow3.4$ (Eqn.~\ref{eqn1} \& Fig.~\ref{fig2}).  RR Lyrae variables, red clump giants, and classical Cepheids provide consistent reddenings for the Galactic bulge, LMC, SMC, and IC 1613 (Table~\ref{reds}).  The $VI$-based RR Lyrae colour-excess relation appears relatively insensitive to metallicity and may be further refined by obtaining multiband photometry for variables in globular clusters \citep[e.g.,][]{sa39,dw77,lay99,cl01,pr03,ho05,ben06,sa09}.  A sizeable portion of the observing program at the Abbey Ridge Observatory \citep{la07,ma08b,tu09c} shall be dedicated to such an endeavour. Modest telescopes may serve a pertinent role in such research \citep{pe80,pe86,sz03,pac06,ge09,tu09c}.    
 
RR Lyrae variables follow $VI$ period-colour and scatter-reduced Wesenheit period-magnitude-colour relations as demonstrated here and elsewhere \citep*{kj97,kw01,so03,di04,ben06,di07,so09,ma09d}.  A pertinent example is the RR Lyrae demographic in the globular cluster IC 4499 (Fig.~\ref{fig11}, photometry from \citealt{wn96}). The Wesenheit function may be inferred without \textit{a priori} knowledge of the colour-excess, and the distances ensue.   Indeed, the Wesenheit function shall be readily employed upon the release of data from the upcoming \textit{Gaia} survey since the relation may be calibrated directly via parallax and apparent magnitudes, mitigating the propagation of uncertainties tied to extinction corrections \citep[\textit{Gaia}:][]{bo03a,ey06,ey09}.   In the interim, however, further work is needed to scrutinize the Wesenheit approach to investigating RR Lyrae variables, and to shift from a broad outlook to assessing finer details (e.g., is the relation marginally non-linear, particularly toward the RRe regime, etc).  Applying a strict [Fe/H]-$M_v$ correlation to infer the distance to individual RR Lyrae variables with differing periods may yield inaccurate results.  The [Fe/H]-$M_v$ relation displays considerable spread at a given metallicity \citep[Fig.~1,][]{pr00}. Abundance estimates often exhibit sizeable random and systemic uncertainties, in contrast to individual pulsation periods. Moreover, the correlation is neither linear or universal \citep[e.g., NGC 6441 \& NGC 6388][]{pr00,bo03b,ca09}.  Applying a strict [Fe/H]-$M_v$ relation to RR Lyrae variables with differing periods at a common zeropoint may yield an acceptable mean distance pending a series of ideal circumstances, including where the overestimated distances for shorter-period variables perfectly balance the underestimated distances of longer period variables \citep[e.g.,][although remedied in \citealt{ma09d} via a reddening-free period-magnitude-colour treatment]{ma09c}.  Admittedly, the aforementioned relation is invaluable in assessing the abundance of a target population to first order, etc.  Yet there are also innumerable advantages to employing Wesenheit and period-magnitude relations to characterize RR Lyrae variables \citep[see also][]{bo03b,da04,da06,da08,ca09}.

Lastly, geometric-based distances to the Galactic center \citep{ei05,re09}, nearby variables \citep{be02,be07,vl07}, open clusters \citep[e.g.,][]{tu02,so05,vl09}, globular clusters \citep[e.g., $\omega$ Cen,][]{vv06}, and the galaxies M33 \& M106 \citep{ar98,br05,he05}: appear to \textit{in sum} bolster and consolidate the scale established by Cepheids and RR Lyrae variables \citep{ma06,sa06,ma08,fe08,fe08b,gr08,sc09,ma09,ma09c,ma09d,tu09}.  Yet a significant challenge remains to identify and then mitigate the uncertainties beyond the $7-10\%$ threshold, beyond first order. 

\subsection*{acknowledgements}
\scriptsize{I am grateful to fellow researchers M. Collinge, T. Sumi, S. Nishiyama, L. Hou, J. Benk{\H o}, A. Walker, J. Nemec, F. Benedict, G. Kopacki, B. Pritzl, A. Layden, A. Dolphin, A. Sarajedini, J. Hartman, I. Soszy{\'n}ski \& the OGLE team, whose comprehensive surveys were the foundation of the study, to the AAVSO (M. Saladyga \& A. Henden), CDS, arXiv, NASA ADS, 2MASS, Hubble Heritage, and the RASC. The following works facilitated the preparation of the research: \citet{el85}, \citet{al98}, \citet{bo03b}, \citet{sm04}, \citet{di04}, \citet{fe08b}, \citet{cl08}, and \citet{mar09}.}

\end{document}